\documentclass{eptcs}
\usepackage[belowskip=-10pt,aboveskip=3pt]{caption}

\usepackage{enumitem}

\usepackage{stfloats}
\usepackage{varwidth}
\usepackage[ruled,vlined]{algorithm2e}

\usepackage{graphicx,color}
\usepackage{mathtools}
\usepackage{amscd}
\usepackage{amsthm}
\usepackage{amsmath}
\usepackage[caption=false]{subfig}
\usepackage{latexsym,amssymb}
\usepackage{enumitem}
\usepackage{tikz}
\usepackage{wrapfig}
\usepackage{xspace}

 \usepackage{multirow}
 \usepackage{pifont}
\usepackage{rotating}

\usepackage{url} 

\newcounter{tempctr}
\newcommand{\breakenumistart}{%
  \setcounter{tempctr}{\value{enumi}}%
  \end{enumerate}%
}
\newcommand{\breakenumiend}{%
  \begin{enumerate}%
  \setcounter{enumi}{\value{tempctr}}%
}

\newtheorem{theorem}{Theorem}[section]

\newtheorem{proposition}[theorem]{Proposition}
\newtheorem{lemma}[theorem]{Lemma}

\theoremstyle{definition}

\newtheorem{semantics}{Semantics}
\newtheorem{example}{Example}

\newcommand{\fig}[1]{Fig.~\ref{fig:#1}}
\newcommand{\figs}[2]{Figs.~\ref{fig:#1} and \ref{fig:#2}}
\newcommand{\tab}[1]{Table~\ref{tab:#1}}
\newcommand{\eq}[1]{(\ref{eq:#1})}

\newcommand{\ex}[1]{Ex.~\ref{ex:#1}}
\newcommand{\secn}[1]{Sect.~\ref{sec:#1}}
\newcommand{\secns}[2]{Sects.~\ref{sec:#1} and \ref{sec:#2}}

\newcommand{\prop}[1]{Prop.~\ref{prop:#1}}

\newcommand{\ie}[1][\ ]{i.e.#1}

\newcommand{\eg}[1][\ ]{e.g.#1}

\newcommand{\cB}{\ensuremath{\mathcal{B}}}
\newcommand{\cC}{\ensuremath{\mathcal{C}}}

\newcommand{\cI}{\ensuremath{\mathcal{I}}}

\newcommand{\sN}{\ensuremath{\mathbb{N}}}

\newcommand{\cT}{\ensuremath{\mathcal{T}}}

\newcommand{\cm}{\ensuremath{\Gamma}}

\SetAlFnt{\scriptsize}
\SetAlCapFnt{\scriptsize}
\SetAlCapNameFnt{\scriptsize}
\SetProcFnt{\scriptsize}
\SetProcNameFnt{\scriptsize}
\SetProcArgFnt{\scriptsize}

\graphicspath{{../graphicsNew/}}

\begin{document}

\title{Architecture Diagrams: A Graphical Language for Architecture Style Specification}

\author{%
Anastasia Mavridou
\qquad
Eduard Baranov
\qquad
Simon Bliudze
\qquad
Joseph Sifakis
\institute{\'Ecole polytechnique f\'ed\'erale de Lausanne, Station 14, 1015 Lausanne, Switzerland}
\email{firstname.lastname@epfl.ch}
}

\def\titlerunning{Architecture Diagrams}
\def\authorrunning{A. Mavridou, E. Baranov, S. Bliudze \& J. Sifakis}

\maketitle
\begin{abstract}
Architecture styles characterise families of architectures sharing common characteristics. We have recently proposed configuration logics for architecture style specification. In this paper, we study a graphical notation to enhance readability and easiness of expression. We study simple architecture diagrams and a more expressive extension, interval architecture diagrams. For each type of diagrams, we present its semantics, a set of necessary and sufficient consistency conditions and a method that allows to characterise compositionally the specified architectures. We provide several examples illustrating the application of the results. We also present a polynomial-time algorithm for checking that a given architecture conforms to the architecture style specified by a diagram.
\end{abstract}

\section{Introduction}
\label{sec:intro}
Software architectures~\cite{perry1992foundations, shaw1996software} describe the high-level structure of a system in terms of components and component interactions. They depict generic coordination principles between types of components and can be considered as generic operators that take as argument a set of components to be coordinated and return a composite component that satisfies by construction a given characteristic property~\cite{AttieBBJS15-architectures-faoc}. 

Many languages have been proposed for architecture description, such as architecture description languages (\eg \cite{medvidovic2000classification, iso2011}), coordination languages (\eg \cite{Papadopoulos1998329, reo}) and configuration languages (\eg \cite{wermelinger2001graph, kramer1990configuration}). All these works rely on the distinction between behaviour of individual components and their coordination in the overall system organization. Informally, architectures are characterized by the structure of the interactions between a set of typed components. The structure is usually specified as a relation, \eg connectors between component ports. 

\begin{figure}[h]
\centering
\includegraphics[scale=0.32]{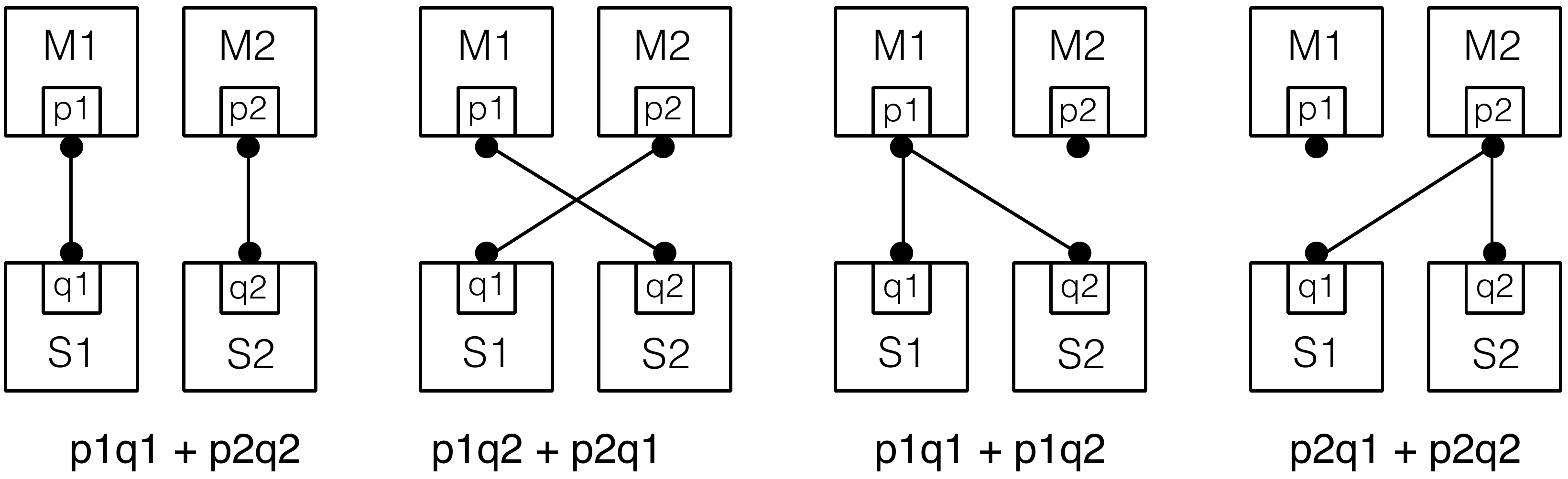}
 \caption{Master/Slave architectures.}
  \label{fig:masterslave}
\end{figure}

\emph{Architecture styles} characterise not a single architecture but a family of architectures sharing common characteristics, such as the types of the involved components and the topology induced by their coordination structure. Simple examples of architecture styles are Pipeline, Ring, Master/Slave, Pipes and Filters. For instance, Master/Slave architectures integrate two types of components, masters and slaves, such that each slave can interact only with one master. \fig{masterslave} depicts four Master/Slave architectures involving two master components $M_1$, $M_2$ and two slave components $S_1$, $S_2$. Their communication ports are respectively $p_1$, $p_2$ and $q_1$, $q_2$. A Master/Slave architecture for two masters and two slaves can be represented as one among the following configurations, \ie sets of connectors:
$\{p_1 q_1, p_2 q_2\}$, $\{p_1 q_2, p_2 q_1\}$, $\{p_1 q_1, p_1 q_2\}$, $\{p_2 q_1, p_2 q_2\}$. A term $p_i q_j$ represents a connector between ports $p_i$ and $q_j$. The four architectures are depicted in \fig{masterslave}. The Master/Slave architecture style denotes all the Master/Slave architectures for arbitrary numbers of masters and slaves.

We have recently proposed configuration logics~\cite{cl-mas} for the description of architecture styles. These are powerset extensions of interaction logics \cite{algconn} 
used to describe architectures. In addition to the operators of the extended logic, they have logical operators on sets of architectures. We have studied higher-order configuration logics and shown that they are a powerful tool for architecture style specification. Nonetheless, their richness in operators and concepts may make their use challenging.

In this paper we explore a different avenue to architecture style specification based on \emph{architecture diagrams}. 
Architecture diagrams describe the structure of a system by showing the system's component types and their attributes for coordination, as well as relationships among component types. Our notation allows the specification of generic coordination mechanisms based on the concept of \emph{connector}. 

Architecture diagrams were mainly developed for architecture style specification in BIP~\cite{AttieBBJS15-architectures-faoc}, where connectors are defined as $n$-ary synchronizations among component ports and do not carry any additional behaviour. Nevertheless, our approach can be extended for architecture style specification in other languages by explicitly associating the required behaviour to connectors.

\begin{figure}[t]
\begin{minipage}[b]{0.5\textwidth}
	\centering
	\includegraphics[scale=0.25]{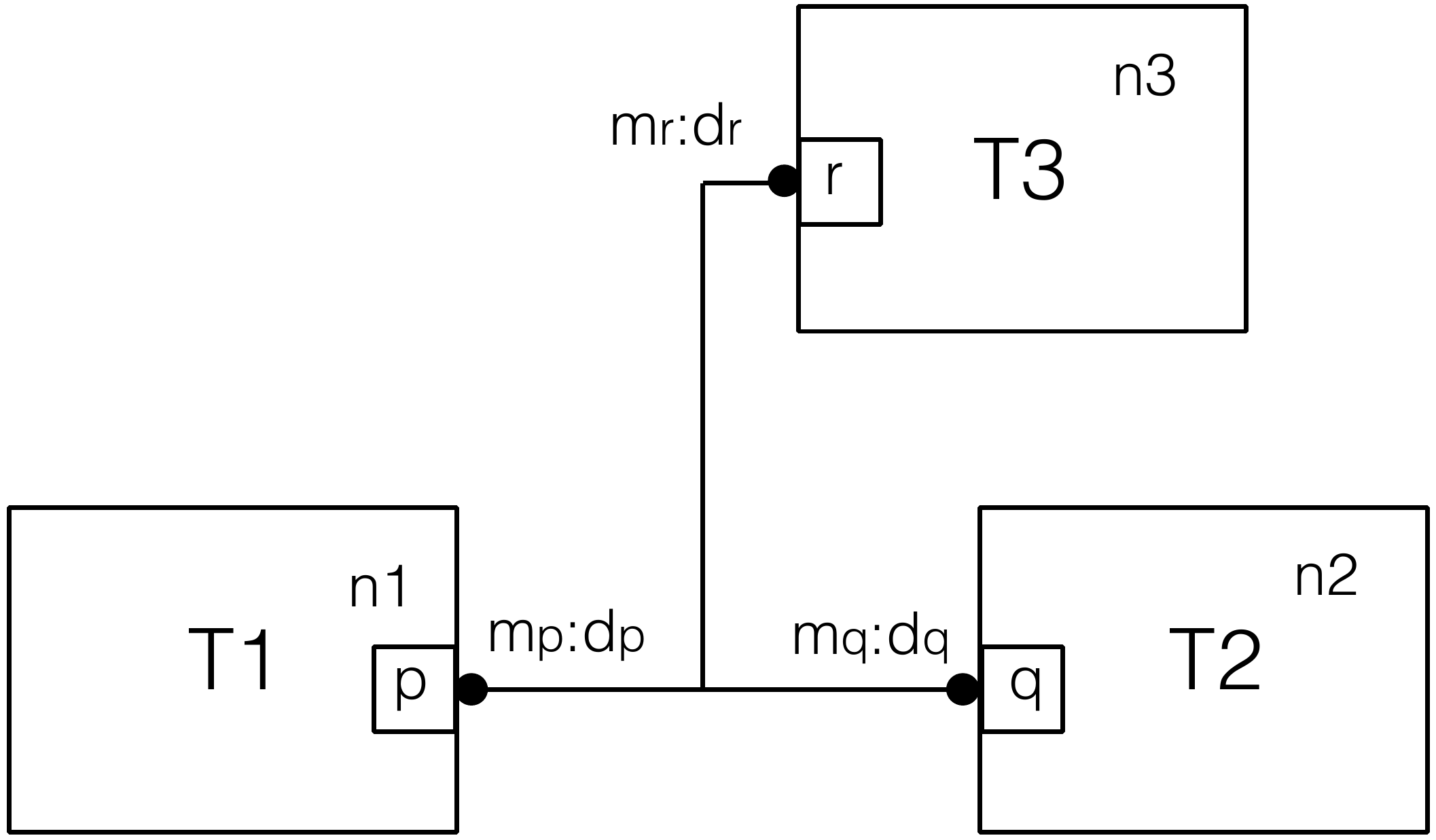}
	\caption{An architecture diagram.}  	
  	\label{fig:ex1}
\end{minipage}
\begin{minipage}[b]{0.5\textwidth}
	\centering
	\includegraphics[scale=0.25]{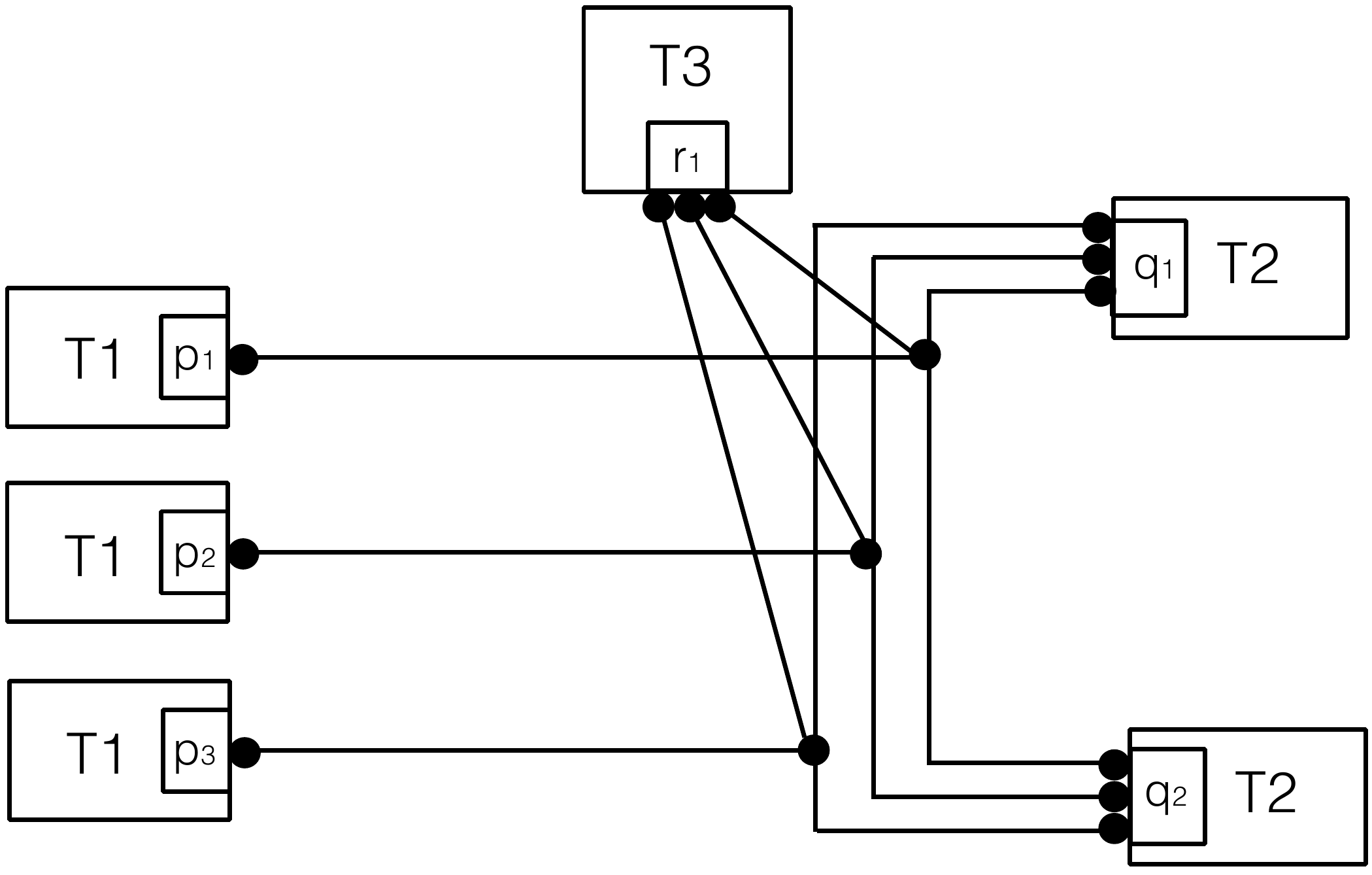}
  	\caption{An architecture.}
  	\label{fig:obtArchEx1}
\end{minipage}
\end{figure}

An architecture diagram consists of a set of \emph{component types}, a \emph{cardinality function} and a set of \emph{connector motifs}. Component types are characterised by sets of \emph{generic ports}. The cardinality function associates each component type with its \emph{cardinality}, \ie number of instances. \fig{ex1} shows an architecture diagram consisting of three component types $T_1$, $T_2$ and $T_3$ with $n_1$, $n_2$ and $n_3$ instances and generic ports $p$, $q$ and $r$, respectively.  Instantiated components have \emph{port instances} $p_i$, $q_j$, $r_k$ for $i,j,k$ belonging to the intervals $[1,n_1]$, $[1,n_2]$, $[1, n_3]$, respectively.

Connector motifs are non-empty sets of generic ports that must interact. Each generic port $p$ in the connector motif has two constraints represented as a pair $m:d$. Multiplicity $m$ is the number of port instances $p_i$ that are involved in each connector. Degree $d$ specifies the number of connectors in which each port instance is involved. The architecture diagram of \fig{ex1} has a single connector motif involving generic ports $p$, $q$ and $r$.

A connector motif  defines a set of possible configurations, where a configuration is a set of connectors.  The meaning of an architecture diagram is a set of architectures that contain the union of all sub-configurations corresponding to each connector motif of the diagram.
\fig{obtArchEx1} shows the unique architecture obtained from the diagram of \fig{ex1} by taking $n_1=3$, $m_p=1$, $d_p=1$; $n_2=2$, $m_q=2$, $d_q=3$, $n_3=1$, $m_r=1$, $d_r=3$. This is the result of composition of constraints for generic ports $p$, $q$ and $r$. For $p$, we have three instances and as both the multiplicity and the degree are equal to $1$, each instance $p_i$ has a single connector lead. For $q$, we have two instances and as the multiplicity is $2$, we have connectors involving $q_1$ and $q_2$ and their total number is equal to $3$ to meet the degree constraint. For $r$, we have a single instance $r_1$ that has three connector leads to satisfy the degree constraint. 

We study a method that allows to characterise compositionally the set of configurations specified by a given connector motif if \emph{consistency conditions} are met. It involves a two-step process. The first step consists in characterising configuration sets meeting the coordination constraints for each generic port $p$ of the connector motif. In the second step, connectors from the sets obtained from step one are fused one by one, so that the multiplicities and the degrees of the ports are preserved, to generate the configuration of the connector motif.

We study two types of architecture diagrams: \emph{simple architecture diagrams} and \emph{interval architecture diagrams}. In the former the cardinality, multiplicity and degree constraints are positive integers, while in the latter they can also be intervals. Interval diagrams are strictly more expressive than simple diagrams. For each type of diagrams we present 1) its syntax and semantics; 2) a set of consistency conditions; 3) a method that allows to characterise compositionally all configurations of a connector motif; 4) examples of architecture style specification. Finally, we present a polynomial-time algorithm for checking that a given diagram conforms to the architecture style specified by a diagram.

A complete presentation, with proofs and additional examples, of the results in this paper can be found in the technical report \cite{MBBS16-Diagrams-TR}. 

The paper is structured as follows. \secns{sad}{iad} present simple and interval architecture diagrams, respectively. \secn{satisfaction} presents an algorithm for checking conformance of diagrams. \secn{relatedwork} discusses related work. \secn{discussion} summarises the results and discusses possible directions for future work.

\vspace{-0.3cm}
\section{Simple Architecture Diagrams}
\label{sec:sad}


\subsection{Syntax and Semantics}

We focus on the specification of generic coordination mechanisms based on the concept of connector. Therefore, the nature and the operational semantics of components are irrelevant.  As in the previous section, we consider that a component interface is defined by its set of ports, which are used for interaction with other components.
Thus, a {\em component type} $T$ has a set of \emph{generic ports} $T.P$.

A {\em simple architecture diagram} $\langle \cT, n, \cC \rangle$ consists of:
1) a set of \emph{component types} $\cT = \{T_1, \dots, T_k\}$;
2) an associated {\em cardinality} function $n : \cT \rightarrow \sN$, where $\sN$ is the set of natural numbers (to simplify the notation, we will abbreviate $n(T_i)$ to $n_i$);
3) a set of \emph{connector motifs} $ \cC = \{\cm_1, \dots, \cm_l\}$ of the form  $\cm =(a, \{m_p:d_p\}_{p \in a})$, where $\emptyset \neq a \subseteq \bigcup_{i=1}^k T_i.P$ is a generic connector and $m_p, d_p \in \sN$ (with $m_p > 0$) are the \emph{multiplicity} and \emph{degree} associated to generic port $p \in a$.

\begin{figure}[t]
\centering
\includegraphics[scale=0.3]{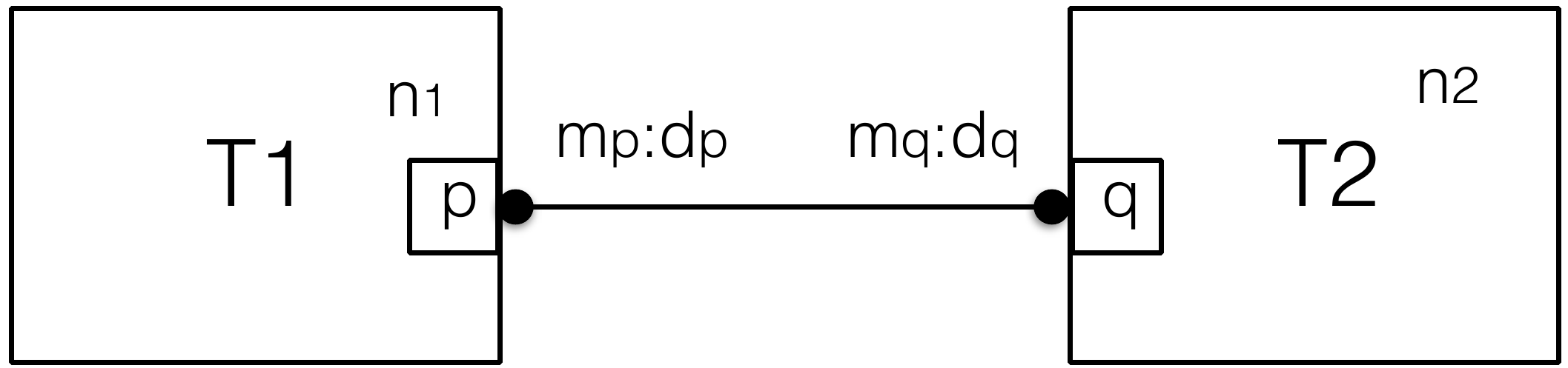}
\caption{A simple architecture diagram.} 

\label{fig:generalSAD}
\end{figure}

\fig{generalSAD} shows the graphical representation of a simple architecture diagram with a connector motif. 
 
An \emph{architecture} is a pair $\langle \cB, \gamma \rangle$, where $\cB$ is a set of components and $\gamma$ is a \emph{configuration}, \ie a set of connectors among the ports of components in $\cB$. We define a connector as a set of ports that must interact. For a component $B \in \cB$ and a component type $T$, we say that {\em $B$ is of type $T$} if the ports of $B$ are in a bijective correspondence with the generic ports in $T$.
Let $B_1,\dots,B_n$ be all the components of type $T$ in $\cB$.  For a generic port $p \in T.P$, we denote the corresponding port instances by $p_1, \dots, p_n$ and its associated cardinality by $n_p = n(T)$.

\begin{semantics}
An architecture $\langle \cB, \gamma \rangle$ \emph{conforms} to a diagram $\langle \cT, n, \cC \rangle$  if, for each $i \in [1,k]$, 
  the number of components of type $T_i$ in $\cB$ is equal to $n_i$ and
  $\gamma$ can be partitioned into disjoint sets $\gamma_1,\dots,\gamma_l$, such that, for each connector motif $\cm_j =(a, \{m_p:d_p\}_{p \in a}) \in \cC$ 
  and each $p \in a$, 
  1)~there are exactly $m_p$ instances of $p$ in each connector in $\gamma_{j}$ and 
2)~each instance of $p$ is involved in exactly $d_p$ connectors in $\gamma_{j}$.
\end{semantics}

We assume that, for any two connector motifs $\Gamma_i = (a, \{m_p^i : d_p^i\}_{p \in a})$ (for $i = 1,2$) with the same set of generic ports $a$, there exists $p \in a$, such that $m_p^1 \neq m_p^2$. 
  Without significant impact on the expressiveness of the formalism,
  this assumption simplifies semantics and analysis. Details are provided in \cite{MBBS16-Diagrams-TR}.

\begin{figure}[h]
\begin{minipage}[b]{0.5\textwidth}
	\centering
	\includegraphics[scale=0.27]{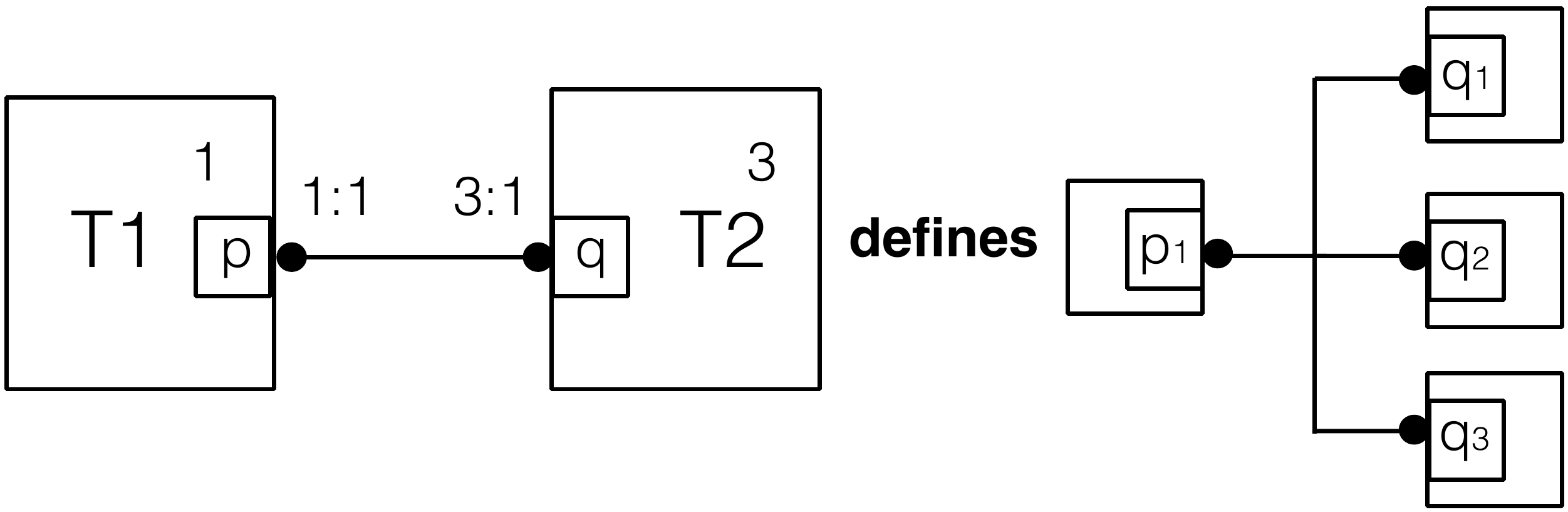}
	\caption{4-ary synchronisation.}  	
  	\label{fig:4arysynch}
\end{minipage}
\begin{minipage}[b]{0.5\textwidth}
	\centering
	\includegraphics[scale=0.27]{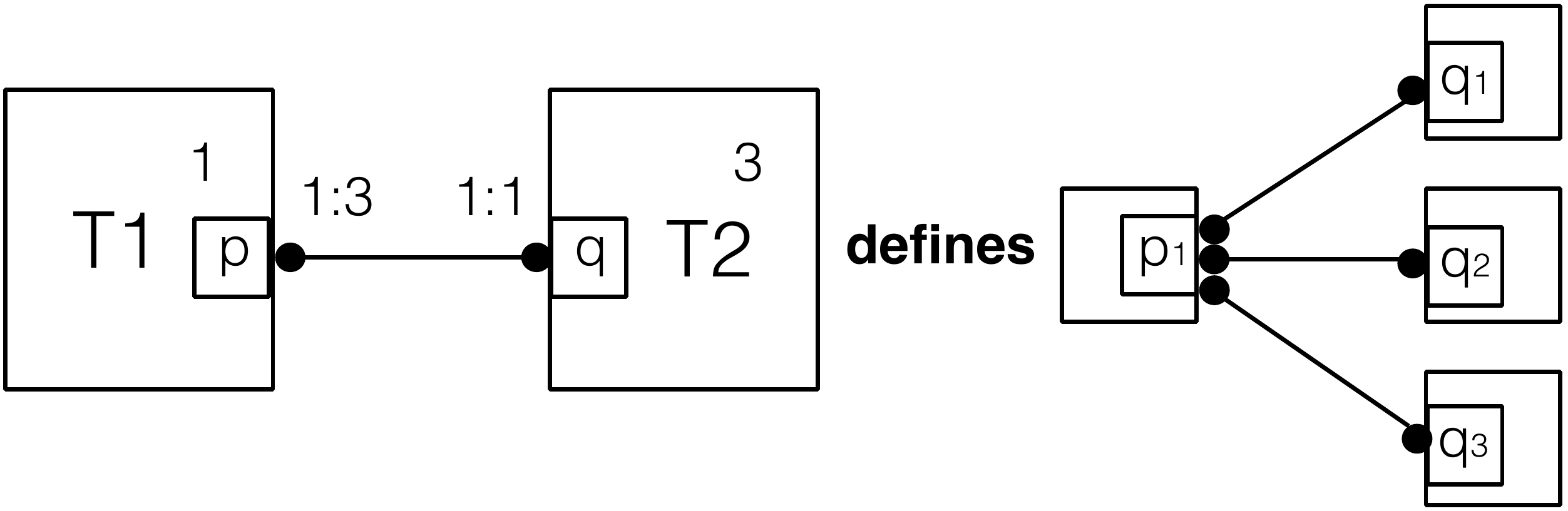}
  	\caption{Binary synchronisation.}
  	\label{fig:binarysynch}
\end{minipage}
\end{figure}

Multiplicity constrains the number of instances of the generic port that must participate in a connector, whereas degree constrains the number of connectors attached to any instance of the generic port. Consider the two diagrams and their conforming architectures shown in \figs{4arysynch}{binarysynch}. They have the same set of component types and cardinalities.  Nevertheless, their multiplicities and degrees differ, resulting in different architectures. 
 
In \fig{4arysynch}, the multiplicity of generic port $p$ is $1$ and the multiplicity of generic port $q$ is $3$, thus, any connector must involve one instance of $p$ and all three instances of $q$. The degree of both generic ports is $1$, so each port instance is involved in exactly one connector. Thus, the diagram defines an architecture with one quaternary connector. 

In \fig{binarysynch} the multiplicities of both generic ports $p$ and $q$ are $1$.  Thus, all connectors are binary and involve one instance of $p$ and one instance of $q$. The degree of $p$ is 3, thus three connectors are attached to each instance. Thus, the diagram defines an architecture with three binary connectors. 

\vspace{-.3cm}
\subsection{Consistency Conditions}
\label{sec:creqs}

Notice that there exist diagrams that do not define any architecture. Let us consider the diagram shown in \fig{generalSAD} with $n_1=3$, $m_p=1$, $d_p=1$, $n_2=2$, $m_q=1$ and $d_q=1$. Since the multiplicity is $1$ for both generic ports $p$ and $q$, a conforming architecture must include only binary connectors involving one instance of $p$ and one instance of $q$. Since the degree of both $p$ and $q$ is $1$, each port instance must be involved in exactly one connector. However, the cardinalities impose that there be three connectors attached to the instances of $p$, but only two connectors attached to the instances of $q$. Both requirements cannot be satisfied simultaneously and thus, no architecture can conform to this diagram.

Consider a connector motif $\cm =(a, \{m_p:d_p\}_{p \in a})$ in a diagram $\langle \cT, n, \cC\rangle$ and a generic port $p \in a$, such that $p \in T.P$, for some $T \in \cT$.  We denote $s_p = n_p\cdot d_p/m_p$ the \emph{matching factor of $p$}.  

A \emph{regular configuration of $p$} is a multiset of connectors, such that 1)~each connector involves $m_p$ instances of $p$ and no other ports and 2)~each of the $n_p$ instances of port $p$ is involved in exactly $d_p$ connectors.
Notice the difference between a configuration and a regular configuration of $p$: the former defines a set of connectors, while the latter defines a multiset of sub-connectors involving only instances of generic port $p$. Considering the diagram in \fig{ex1} and the architecture in \fig{obtArchEx1} the only regular configuration of $r$ is the multiset $\{r_1,\ r_1,\ r_1\}$.  The three copies of the singleton sub-connector $r_1$ are then fused with sub-connectors $p_i q_1 q_2$ ($i = 1,2,3$), resulting in a configuration with three distinct connectors.

\begin{lemma}
 Each regular configuration of a port $p$ has exactly $s_p$ connectors. 
\end{lemma}

\prop{srequirement} provides the necessary and sufficient conditions for a simple architecture diagram to be consistent, \ie to have at least one conforming architecture. 
The multiplicity of a generic port must not exceed the number of component instances that contain this port. The matching factors of all ports participating in the same connector motif must be equal integers.  Finally, since the number of distinct connectors of a connector motif is bounded and equal to $\prod_{q \in a} {\binom{n_q}{m_q}}$, there must be enough connectors to build a configuration.  
Since, by the semantics of diagrams, connector motifs correspond to disjoint sets of connectors, these conditions are applied separately to each connector motif. 

\begin{proposition}
\label{prop:srequirement}
A simple architecture diagram has a conforming architecture iff, for each connector motif $\cm = (a, \{m_p:d_p\}_{p \in a})$ and each $p \in a$, we have:
1)~$m_p \le n_p$; 
2)~$\forall q \in a,\ s_p = s_q \in \mathbb{N}$ and
3)~$s_p \le \prod_{q \in a} {\binom{n_q}{m_q}}$.
\end{proposition}


\subsection{Synthesis of Configurations}
\label{sec:ssynth}

The synthesis procedure for each connector motif has the following two steps: 1) we find regular configurations for each generic port; 2) we fuse these regular configurations generating global configurations specified by the connector motif. 

\subsubsection{Regular Configurations of a Generic Port}
\label{sec:rconfp}
We start with an example illustrating the first step of the synthesis procedure for a port $p$.

\begin{figure} [h]
  \centering
  \includegraphics[width=\textwidth]{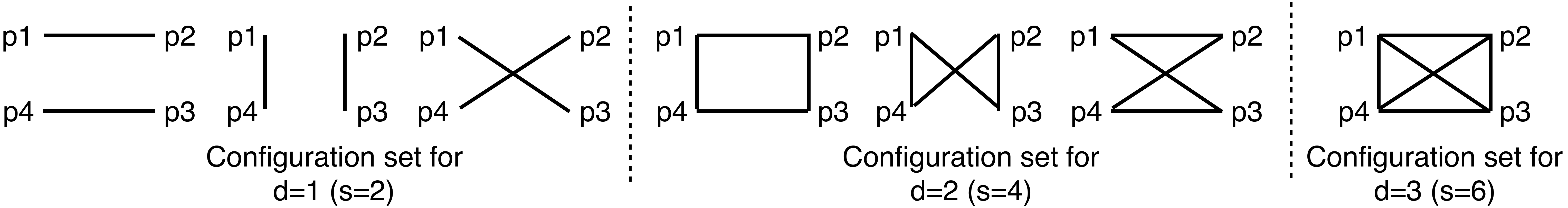}
  \caption{Regular configurations of $p$ with $n_p=4$, $m_p=2$.}
  \label{fig:rconf}
\end{figure}

\begin{example}
\label{ex:rconf}
Consider a port $p$ with $n_p=4$ and $m_p=2$. There are $6$ connectors of multiplicity $2$: $p_1p_2$, $p_1p_3$, $p_1p_4$, $p_2p_3$, $p_2p_4$, $p_3p_4$, which correspond to the set of edges of a complete graph with vertices $p_1$, $p_2$, $p_3$, $p_4$. The regular configurations of $p$ for $d_p=1,2,3$, where each edge appears at most once are shown in \fig{rconf}. 
\end{example}

We provide an equational characterisation of all the regular configurations (\ie multisets of connectors) of a generic port $p$.  Given $n_p$, $m_p$, $d_p$,
for port instances $p_1, \dots, p_{n_p}$, we associate a column vector of non-negative integer variables $X=[x_1,\dots ,x_w]^T$ to the set $\{a_i\}_{i \in [1,w]}$ of different connectors, where $w= \binom{n_p}{m_p}$. 

Consider \ex{rconf} and variables $x_1, \dots, x_6$ representing the number of occurrences in a regular configuration of the connectors $p_1p_2,\, p_1p_3,\, p_1p_4,\, p_2p_3,\, p_2p_4,\,$ $p_3p_4$, respectively.  All the regular configurations, for $d_p = 1,2,3$, represented as vectors of the form $[x_1,\dots, x_6]$
are listed in \tab{vectors}. Notice that vectors for $d_p > 1$ can be obtained as linear combinations of the vectors for $d_p = 1$.

\begin{table} [t]
\centering
\caption{Vector representation of regular configurations.}
\label{tab:vectors}
\arraycolsep=3pt
$\begin{array}{c|cc|cccc}
  \hline
  d_p = 1
  & \multicolumn{2}{c|}{d_p = 2}
  & \multicolumn{4}{c}{d_p = 3}
  \\\hline
  [100001] & [110011] & [200002] & [111111] & [120021] & [210012] & [300003]
  \\{}
  [010010] & [101101] & [020020] &          & [012210] & [021120] & [030030]  
  \\{}
  [001100] & [011110] & [002200] &          & [102201] & [201102] & [003300] 
  \\\hline
\end{array}$
\end{table}

For $p$, we define an $n_p \times w$ incidence matrix $G = [g_{i,j}]_{n_p \times w}$ with $g_{i,j}= 1$ if  $p_i \in a_j$ and $g_{i,j} = 0$ otherwise.
We have $GX=D$, where $D=[d_p, \dots, d_p]$ ($d_p$ repeated $n_p$ times).  Any non-negative integer solution of this equation defines a regular configuration of $p$.
For \ex{rconf}, the equations are:
\begin{equation}
  \label{eq:regconf}
  \begin{cases} 
    x_1+x_2+x_3=d_p\,,\\
    x_1+x_4+x_5=d_p\,,\\
    x_2+x_4+x_6=d_p\,,\\
    x_3+x_5+x_6=d_p\,,
  \end{cases}
  \text{which is equivalent to\quad}
  \begin{cases}
    x_1+x_2+x_3=d_p\,,\\
    x_3=x_4\,,\\
    x_2=x_5\,,\\
    x_1=x_6\,.
  \end{cases}
\end{equation}
Notice that the vectors of \tab{vectors} are solutions of \eq{regconf}. 

\vspace{-0.3cm}
\subsubsection{Configurations of a Connector Motif}
\label{sec:confMotif}

Let $\cm = (a, \{m_p:d_p\}_{p\in a})$ be a connector motif such that 
all generic ports of $a = \{p^1,\dots,p^v\}$ have the same integer matching factor $s$.  For each $p^j \in a$, let $\gamma^j = \{a_i^j\}_{i \in [1,s]}$ 
be a regular configuration of $p^j$
. For arbitrary permutations $\pi_j$ 
of $[1,s]$, a set $\{a_i^1 \cup \bigcup_{j=2}^v a_{\pi_j(i)}^j\}_{i \in [1,s]}$ 
is a configuration specified by the connector motif.

In order to provide an equational characterisation of the connector motif, we consider, for each $j \in [1,v]$, a corresponding solution vector $X^j$ 
of equations $G^j X^j = D^j$ 
characterising the regular configurations of $p^j$. 
We denote by $w^j$ 
the dimension of the vector $X^j$.

In order to characterise the configurations of connectors conforming to $\cm$, we consider, for each configuration, the $v$-dimensional matrix $E= [e_{i_1,\dots,i_v}]_{w^1 \times \dots \times w^v}$ of 0-1 variables, such that $e_{i_1,\dots,i_v} = 1$ if the connector $a_{i_1}^1 \cup \dots \cup a_{i_v}^v$ belongs to the configuration and $0$ otherwise.  By definition, the sum of all elements in $E$ is equal to $s$.  Moreover, the following equations hold:
\begin{equation}
  \label{eq:confeq}
  \left\{
  \renewcommand{\arraystretch}{1.2}
  \begin{array}{r@{\ }c@{\ }ll}
    x_i^1&=&\Sigma_{i_2,i_3\dots,i_v}\ e_{i,i_2,\dots,i_v}\,,
    &\text{for $i \in [1,w^1]$,}
    \\
    x_i^2&=&\Sigma_{i_1,i_3,\dots,i_v}\ e_{i_1,i,\dots,i_v}\,,
    &\text{for $i \in [1,w^2]$,}
    \\
    &\vdots
    \\
    x_i^v&=&\Sigma_{i_1,i_2,\dots,i_{v-1}}\ e_{i_1,\dots,i_{v-1},i}\,,
    &\text{for $i \in [1,w^v]$.}
  \end{array}
  \right.
\end{equation}
For instance, for a fixed $i\in [1,w^1]$, $e_{i,i_2,\dots,i_v}$ describe all connectors that contain $a_i^1$. The regular configuration $\gamma^1$ is characterised by $X^1$, enforcing that $a_i^1$ belongs to $x_i^1$ connectors.  
The set of linear equations \eq{confeq}, combined with the sets of linear equations $G^j X^j = D^j$, for $j \in [1,v]$, fully characterises the configurations of $\cm$ and can be used to synthesise architectures from architecture diagrams.

\begin{example}
\label{ex:mergerconf}
Consider a diagram  $\bigl(\{T_1,T_2\}, n, \{\cm\}\bigr)$, where $T_1 = \{p\}$,  $T_2 = \{q\}$, $n(T_1) = n(T_2) = 4$ and $\cm = (pq, \{(m_p:d_p, m_q:d_q)\})$ with $m_p=2$, $m_q=3$.  The corresponding equations $G_p X = D_p$, $G_q Y = D_q$ can be rewritten as
\begin{equation}
  \label{eq:mergerconf}
  \begin{cases}
    x_1+x_2+x_3=d_p,\\
    x_3=x_4, \
    x_2=x_5, \
    x_1=x_6, \
  \end{cases} 
  \text{and\quad} 
  \begin{cases}
    3 y_1=d_q,\\
    y_1=y_2=y_3=y_4.
  \end{cases}
\end{equation}
Together with the constraints $x_i = \Sigma_j e_{i,j}$ and $y_j = \Sigma_i e_{i,j}$, for $E=[e_{i,j}]_{6\times 4}$, equations \eq{mergerconf} completely characterise all the configurations conforming to $\cm$.
\end{example}

The same methodology can be used to synthesise configurations with additional constraints.  To impose that some specific connectors must be included, whereas other specific connectors must be excluded from the configurations, the corresponding variables in the matrix $E$ are given fixed values: 1 (resp. 0) if the connector must be included (resp. excluded) from the configurations.  The rest of the synthesis procedure remains the same. 
\begin{example}
\label{ex:synth}
Let us consider the diagram shown in \fig{generalSAD} with $n_1=4$, $m_p=2$, $d_p=2$, $n_2=4$, $m_q=3$ and $d_q=3$.
We want to synthesise the configurations of this diagram with the following additional constraints: connectors $p_1p_2q_1q_2q_3$ and $p_1p_3q_2q_3q_4$ must be included, whereas connector $p_2p_4q_1q_2q_4$ must be excluded.

First, we compute the vectors $X$ and $Y$ that represent the regular configurations of generic ports $p$ and $q$, respectively.  Variables $x_1, \dots, x_6$ represent the number of occurrences in a configuration of the connectors $p_1p_2$, $p_1p_3$, $p_1p_4$, $p_2p_3$, $p_2p_4$, $p_3p_4$, respectively.  Variables $y_1, \dots, y_4$ represent the number of occurrences in a configuration of the connectors $q_1 q_2 q_3$, $q_1 q_2 q_4$, $q_1 q_3 q_4$, $q_2 q_3 q_4$, respectively.

Vector $X$ can take one of the following values for $d_p = 2$: $[110011]$, $[101101]$, $[011110]$, $[200002]$, $[020020]$ or $[002200]$ (\ex{rconf}). Regular configurations of $q$ are characterised by the 
equations $3y_1=d$ and $y_1=y_2=y_3=y_4$ (\ex{mergerconf}).
For $d = 3$ there is a single solution $Y = [1111]$.

We now consider the matrix $E$, where we fix $e_{1,1} = e_{2,4} = 1$ and $e_{5,2} = 0$ to impose the additional synthesis constraints as shown in \fig{matrix}.
Since, for all $i \in [1,6]$, we have $x_i = \Sigma_j\, e_{i,j}$ and, for all $j \in [1,4]$, we have $y_j = \Sigma_i\, e_{i,j}$, we can deduce that the only possible valuation for $X$, $Y$ and $E$ is the one shown in \fig{valuation}
corresponding to configuration
$\{p_1p_2q_1q_2q_3,\ p_1p_3q_2q_3q_4,\
p_2p_4q_1q_3q_4,\ p_3p_4q_1q_2q_4\}.$
\end{example}

\begin{figure} 
\begin{minipage}[b]{0.5\textwidth}
	\centering
\[
E = \bordermatrix{ ~ & y_1 & y_2 & y_3 & y_4 \cr
x_1 & 1       & e_{1,2} & e_{1,3} & e_{1,4} \cr
x_2 & e_{2,1} & e_{2,2} & e_{2,3} & 1 \cr
x_3 & e_{3,1} & e_{3,2} & e_{3,3} & e_{3,4} \cr
x_4 & e_{4,1} & e_{4,2} & e_{4,3} & e_{4,4} \cr
x_5 & e_{5,1} & 0       & e_{5,3} & e_{5,4} \cr
x_6 & e_{6,1} & e_{6,2} & e_{6,3} & e_{6,4} \cr	
}
\]
	\caption{Matrix $E$ with fixed values.}  	
  	\label{fig:matrix}
\end{minipage}
\begin{minipage}[b]{0.5\textwidth}
	\centering
\[
E \quad = \quad
\bordermatrix{ {}_X\!\diagdown{\scriptstyle Y} & 1 & 1 & 1 & 1 \cr
1 & 1 & 0 & 0 & 0 \cr
1 & 0 & 0 & 0 & 1 \cr
0 & 0 & 0 & 0 & 0 \cr
0 & 0 & 0 & 0 & 0 \cr
1 & 0 & 0 & 1 & 0 \cr
1 & 0 & 1 & 0 & 0 \cr	
}
\]
  	\caption{Valuation of matrix $E$.}
  	\label{fig:valuation}
\end{minipage}
\end{figure}

\vspace{-0.3cm}
\subsection{Architecture Style Specification Examples}
\label{sec:SADarchstyle}

\begin{figure} [ht]
\centering
\includegraphics[scale=0.2]{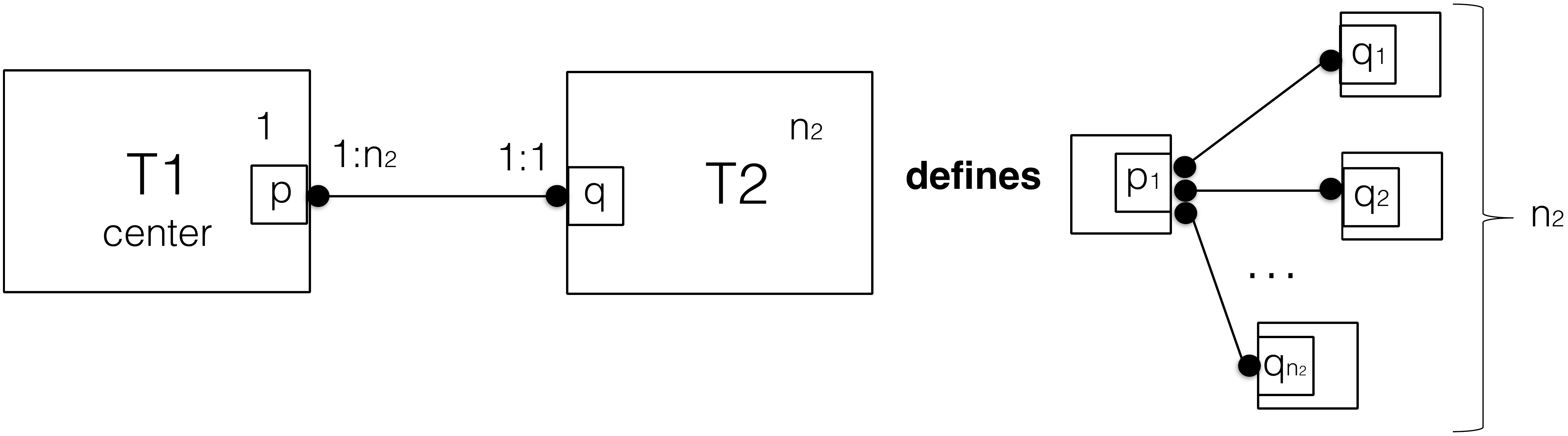}
\caption{Star architecture style.}
\label{fig:starstyle}
\end{figure}

\begin{example}
\label{ex:starstyle}
The Star architecture style consists of a single center component of type $T_1 = \{p\}$ and $n_2$ components of type $T_2 = \{q\}$. 
The central component is connected to every other component by a binary connector and there are no other connectors. 
The diagram in \fig{starstyle} graphically describes this style.
\end{example}

\begin{figure} [ht]
\centering
\includegraphics[scale=0.2]{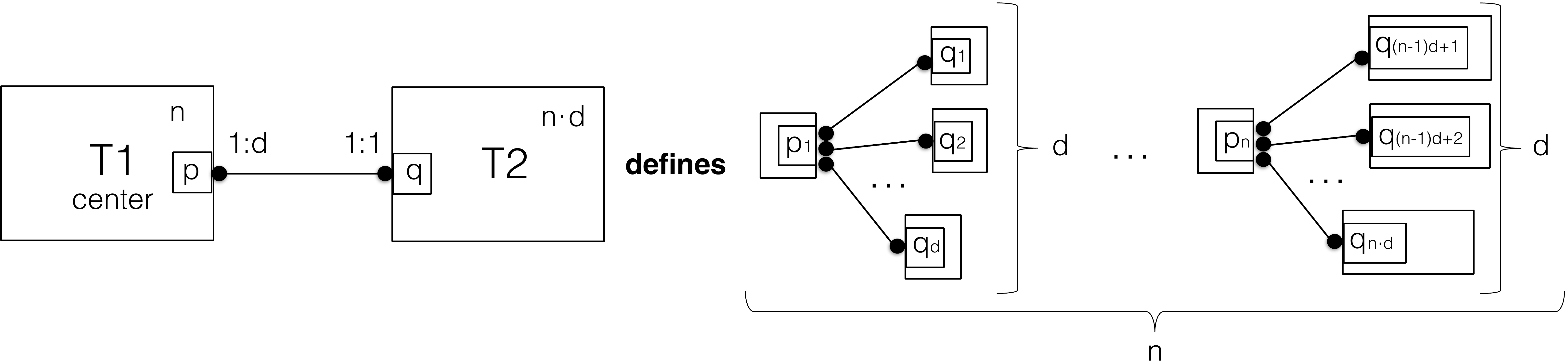}
\caption{Multi-star architecture style.}
\label{fig:nstarstyle}
\end{figure}

\begin{example}
We now consider the multi-star extension of the Star architecture style,  with $n$ center components of type $T_1$, each connected to $d$ components of type $T_2$ by binary connectors.  As in \ex{starstyle}, there are no other connectors. 
The diagram of \fig{nstarstyle} graphically describes this architecture style.
\end{example}

\vspace{-0.5cm}


\section{Interval Architecture Diagrams}
\label{sec:iad}

To enhance the expressiveness of diagrams we introduce interval architecture diagrams where the cardinalities, multiplicities and degrees can be intervals. With simple architecture diagrams we cannot express properties such as \emph{``component instances of type $T$ are optional''}. Let us consider the example of \fig{masterslave} that shows four Master/Slave architectures involving two masters and two slaves. In this example, one of the masters might be optional, \ie it might not interact with any slaves.  In the first two architectures of \fig{masterslave} each master interacts with one slave, however, in the last two architectures one master interacts with both slaves while the other master does interacts with no slaves. In other words, the degree of $m$ varies from $0$ to $2$ and cannot be represented by an integer. 
 
\vspace{-0.3cm}
\subsection{Syntax and Semantics}

An {\em interval architecture diagram} $\langle \cT, n, \cC\rangle$ consists of:
1) a set of \emph{component types} $\cT = \{T_1, \dots, T_k\}$; 2) a {\em cardinality} function $n : \cT \rightarrow \sN^2$, associating, to each $T_i \in \cT$, an interval $n(T_i) = [n_i^l, n_i^u] \subseteq \sN$ (thus, $n_i^l \leq n_i^u$); 3) a set of \emph{connector motifs} $ \cC = \{\cm_1, \dots, \cm_l\}$ of the form $\cm =\Big(a, \{ty[m_p^l,m_p^u]: ty[d_p^l,d_p^u]\}_{p \in a}\Big)\,,$
where $\emptyset \neq a \subseteq \bigcup_{i=1}^k T_i.P$ is a generic
    connector and $ty[m_p^l,m_p^u], ty[d_p^l,d_p^u]$, with $[m_p^l,m_p^u],[d_p^l,d_p^u] \subseteq \sN$ non-empty intervals and $ty \in \{mc, sc\}$ ($mc$ means ``multiple    choice'', whereas $sc$ means ``single choice''), are, respectively, \emph{multiplicity} and \emph{degree} constraints associated to $p \in a$,

\begin{semantics}
An architecture $\langle \cB, \gamma \rangle$ \emph{conforms} to an interval architecture diagram $\langle \cT, n, \cC\rangle$ if, for each $i \in [1,k]$, the number of components of type $T_i$ in $\cB$ lies in $[n_i^l,n_i^u]$ and $\gamma$ can be partitioned into disjoint sets $\gamma_1,\dots,\gamma_l$, such that for each connector motif $\cm_j = \bigl(a, \{ty[m_p^l,m_p^u]: ty[d_p^l,d_p^u]\}_{p \in a}\bigr) \in \cC$ 
and each $p \in a$: 1) there are $m_p \in [m_p^l,m_p^u]$ instances of $p$ in each connector in $\gamma_{j}$; in case of a single choice interval the number of instances of $p$ is equal in all connectors in $\gamma_{j}$; 2) each instance of $p$ is involved in $d_p \in [d_p^l,d_p^u]$ connectors in $\gamma_{j}$; in case of a single choice interval, the number of connectors involving an instance of $p$ is the same for all instances of $p$.
\end{semantics}

In other words, each generic port $p$ has an associated pair of intervals defining its multiplicity and degree. The interval attributes specify whether these constraints are uniformly applied or not. We write $sc[x,y]$ (single choice) to mean that the same multiplicity or degree is applied to each port instance of $p$.  We write $mc[x,y]$ (multiple choice) to mean that different multiplicities or degrees can be applied to different port instances of $p$, provided they lie in the interval.

We assume that, for any two connector motifs 
$\Gamma_i = (a,\allowbreak \{ty[m_p^l,m_p^u]_i: ty[d_p^l,d_p^u]_i\}_{p \in a})\,
\text{ for } i \in \{1,2\},
$
with the same set of generic ports $a$, there exists $p \in a$ such that $[m_p^l,m_p^u]_1 \cap [m_p^l,m_p^u]_2 = \emptyset$.  Similarly to simple architecture diagrams, without significant impact on the expressiveness of the formalism,  this assumption greatly simplifies semantics and analysis.

\begin{figure} [t]
\centering
\includegraphics[scale=0.3]{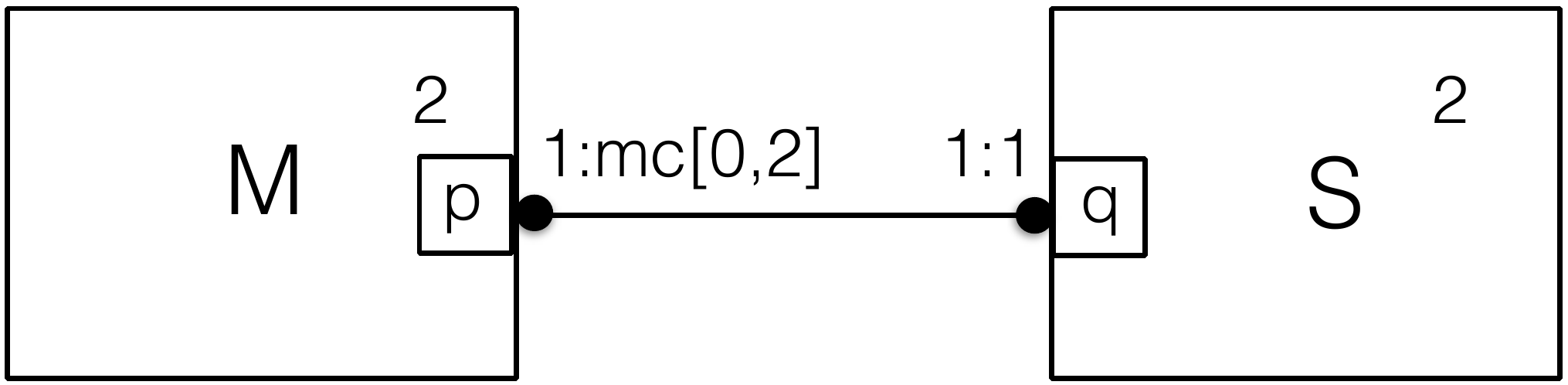}
\caption{Architecture diagram for architectures in \fig{masterslave}.}
\label{fig:mastslav}
\end{figure}
\begin{example}
\label{ex:ims}
The diagram in \fig{mastslav}
defines the set of architectures shown in \fig{masterslave}. Notice that the degree of generic port $p$ is the multiple choice interval $[0,2]$, since one master component may be connected to two slaves, while the other master may have no connections. For the sake of simplicity, we represent intervals $[x,x]$, $mc[x,x]$ and $sc[x,x]$ as $x$.
\end{example}

\begin{proposition}
Interval architecture diagrams are strict\-ly more expressive than simple architecture diagrams.
\end{proposition}

\vspace{-0.3cm}
\subsection{Consistency Conditions}

Similarly to simple diagrams, there are interval diagrams that do not define any architectures. \prop{requirement} provides the necessary and sufficient conditions for the consistency of interval diagrams. 
A connector cannot contain more port instances than there exist in the system. Thus, the lower bound of multiplicity should not exceed the maximal number of instances of the associated component type. For all generic ports of a connector motif, there should exist a common matching factor that does not exceed the maximum number of different connectors between these ports.
These conditions are a generalisation of \prop{srequirement}.

To simplify the presentation we use the following notion of choice
function.  Let $\cI_T$ and $\cI$ be the sets of, respectively, typed
intervals and intervals, as in the definition of interval diagrams
above.  A function $g: \cI_T \rightarrow \cI$ is a {\em choice
  function} if it satisfies the following constraints:
\[
g(ty[x,y]) = 
\begin{cases}
  [x,y], & \text{if $ty = mc$,}\\
  [z,z], & \text{for some $z \in [x,y]$, if $ty = sc$.}
\end{cases}
\]

\vspace{-0.3cm}

\label{sec:cintervalreqs}
\begin{proposition}
\label{prop:requirement}
An interval architecture diagram $\langle\cT,\allowbreak n,\allowbreak \cC\rangle$ is consistent iff, for each $T \in \cT$, there exists a cardinality $n_i \in [n_i^l,n_i^u]$ and, for each connector motif $(a, \{M_p: D_p\}_{p \in a}) \in \cC$ and each $p \in a$, there exist choice functions $g_p^m, g_p^d$, such that, for $[m_p^l,m_p^u] = g_p^m(M_p)$ and $[d_p^l,d_p^u] = g_p^d(D_p)$ hold: 

\begin{enumerate}[topsep=1pt,itemsep=-3pt,partopsep=1ex,parsep=1ex]
\item $m_p^l \leq n_p$, for all $p \in a$, (where $n_p = n_i$ for $p \in T_i.P$),
\item $U \cap \bigcap_{p \in a} s_p \neq \emptyset$, where 
  $U = \bigl[1, \prod_{p \in a} \sum_{m = m_p^l}^{m_p^u} \binom{n_p}{m}\bigr]$, 
  $\text{and}\\
  s_p = 
  \begin{cases} 
    \left[\frac{n_p \cdot d_p^l}{m_p^u}, \frac{n_p \cdot d_p^u}{m_p^l} \right] 
    \cap \sN\,,
    &\text{ if }  m_p^l > 0,
    \\[8pt]
    \left[\frac{n_p \cdot d_p^l}{m_p^u}, \infty\right) \cap \sN\,,
      &\text{ if } m_p^l = 0.
  \end{cases}$
\end{enumerate}
\end{proposition}

\vspace{-0.3cm}
\subsection{Synthesis of Configurations}

The equational characterisation in \secn{ssynth} can be generalised, using systems of inequalities with some additional variables, to interval architecture diagrams.
Below, we show how to characterise the configurations induced by $n$ instances of a generic port $p$ with the associated degree interval $ty[d_p^l,d_p^u]$.

For a given multiplicity $m$, let $X=[x_1, \dots,x_w]^T$ be the column vector of integer variables, corresponding to the set $\{a_i\}_{i \in [1,w]}$ (with $w = \binom{n_p}{m_p}$) of connectors of multiplicity $m$, involving port instances $p_1, \dots, p_n$.  
Let $G$ be the incidence matrix $G = [g_{i,j}]_{n \times w}$ with $g_{i,j} = 1$ if $p_i \in a_j$ and $g_{i,j} = 0$ otherwise. 
The configurations induced by the $n$ instances of $p$ are characterised by 
the equation $GX=D$, where $D = [d_1,\dots,d_n]^T$ and the additional (in)equalities:
\begin{equation}
  \label{eq:dinterval}
  \begin{aligned}
    &d_1= \dots =d_n=d\text{ and }d_p^l \leq d \leq d_p^u, 
    &&\text{for }ty=sc,\\
    &d_p^l \leq d_1 \leq d_p^u,\dots, d_p^l \leq d_n \leq d_p^u, 
    &&\text{for }ty=mc. 
  \end{aligned}
\end{equation}

\begin{example}
As in \ex{rconf}, consider a generic port $p$ and $n_p=4$, $m_p=2$.
For the degree interval $sc[1,3]$, the corresponding constraints are
$1 \leq d \leq 3$, $x_1+x_2+x_3 = d$, $x_4 = x_3$, $x_5 = x_2$, $x_6 = x_1$.
For the degree interval $mc[1,3]$ the corresponding constraints are $1 \leq d_i \leq 3$, for $i \in [1,4]$, $x_1+x_2+x_3 = d_1$, $x_1+x_4+x_5 = d_2$, $x_2+x_4+x_6 = d_3$, $x_3+x_5+x_6 = d_4$.
\end{example}

Suppose that the multiplicity of $p$ in the motif is given by an
interval $ty[m_p^l,m_p^u]$.  Contrary to the degree, multiplicity does
not appear explicitly as a variable in the constraints.  Instead, it
influences the number and nature of elements in both the matrix
$G$ and vector $X$.
Therefore, for single choice (\ie $ty = sc$), the configurations induced by $n$ instances of $p$ are characterised by the disjunction of the instantiations of the system of equalities combining $G_m X_m = D$ with \eq{dinterval}, for $m \in [m_p^l,m_p^u]$.
For multiple choice (\ie $ty = mc$), all the configurations are 
characterised by the system combining \eq{dinterval} with
$
\sum_{m\in[m_p^l,m_p^u]} (G_m X_m) = D\,.
$ 

Notice that the above modifications for interval-defined
multiplicity are orthogonal to those in \eq{dinterval}, accommodating
for interval-defined degree.
Similarly to the single-choice case for multiplicity, for
interval-defined cardinality, the configurations are characterised by
taking the disjunction of the characterisations for all values $n \in
[n^l,n^u]$.
Based on the above characterisation for the configurations of one
generic port, global configurations can be characterised by systems of
linear constraints in the same manner as for simple architecture
diagrams.
\vspace{-0.3cm}

\subsection{Architecture Style Specification Examples}

\begin{figure} [t]
  \centering
  \includegraphics[scale=0.32]{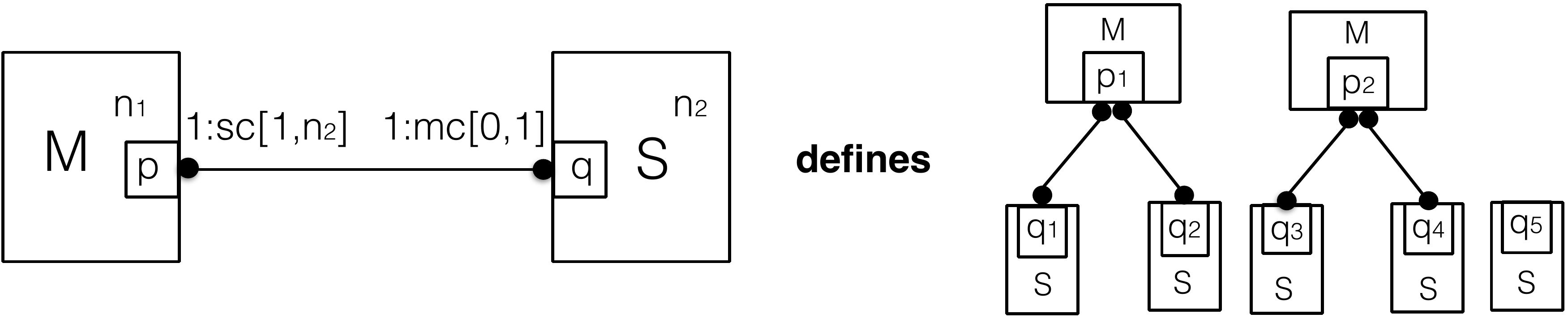}
  \caption{Master/Slave architecture style and conforming architecture.}
  \label{fig:imasterslave}
\end{figure}

\begin{example}
\label{ex:imasterslave}
The diagram of \fig{imasterslave} describes a particular Master/Slave architecture style and a conforming architecture for $n_1=2$ and $n_2=5$.

We require that each slave interact with at most one master and that each master be connected to the same number of slaves.  
Multiplicities of both generic ports $p$ and $q$ are equal to $1$, allowing only binary connectors between a master and a slave. The single choice degree of generic port $p$ ensures that all port instances are connected to the same number of connectors which is a number in $[1,n_2]$. The multiple choice degree of generic port $q$ ensures that all port instances are connected to at most one master. 
\end{example}

\begin{example}
\begin{figure} [ht]
\centering
\includegraphics[scale=0.23] {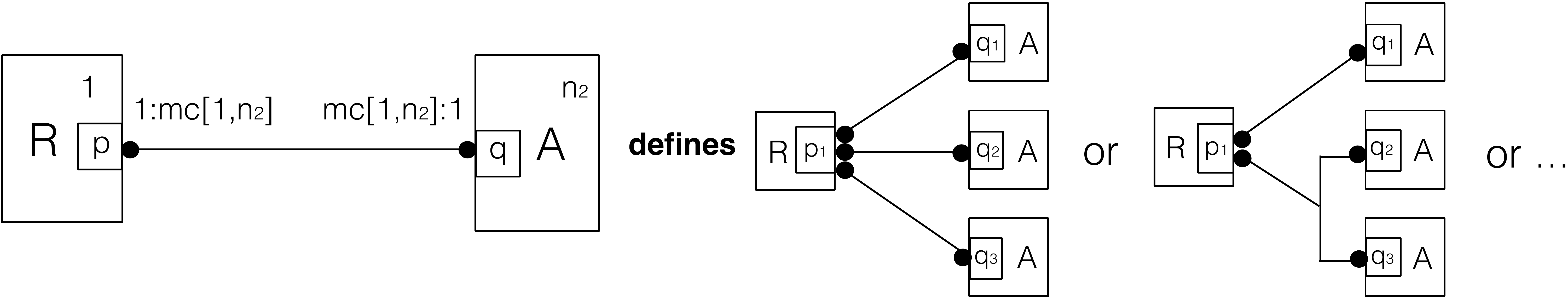}
 \caption{Repository architecture style and conforming architectures.}
  \label{fig:repStyle}
\end{figure}

The diagram in the left of \fig{repStyle} describes the Repository architecture style involving a single instance of a component of type R and an arbitrary number $n_2$ of data-accessor components of type $A$. We require that all connectors involve the $R$ component. In the right of \fig{repStyle}, we show conforming architectures for $n_2=3$. 
\end{example}

\begin{example}
The Map-Reduce architecture style~\cite{mapreduce}
 allows processing large data-sets, such as those found in search engines and social networking sites.  \fig{mapsstyle} graphically describes the Map-Reduce architecture style.  A conforming architecture for $n_1=3$ and $n_2=2$ is shown in \fig{mapsconf}. 

\begin{figure}[t]
\begin{minipage}[b]{0.5\textwidth}
	\centering
	\includegraphics[scale=0.25]{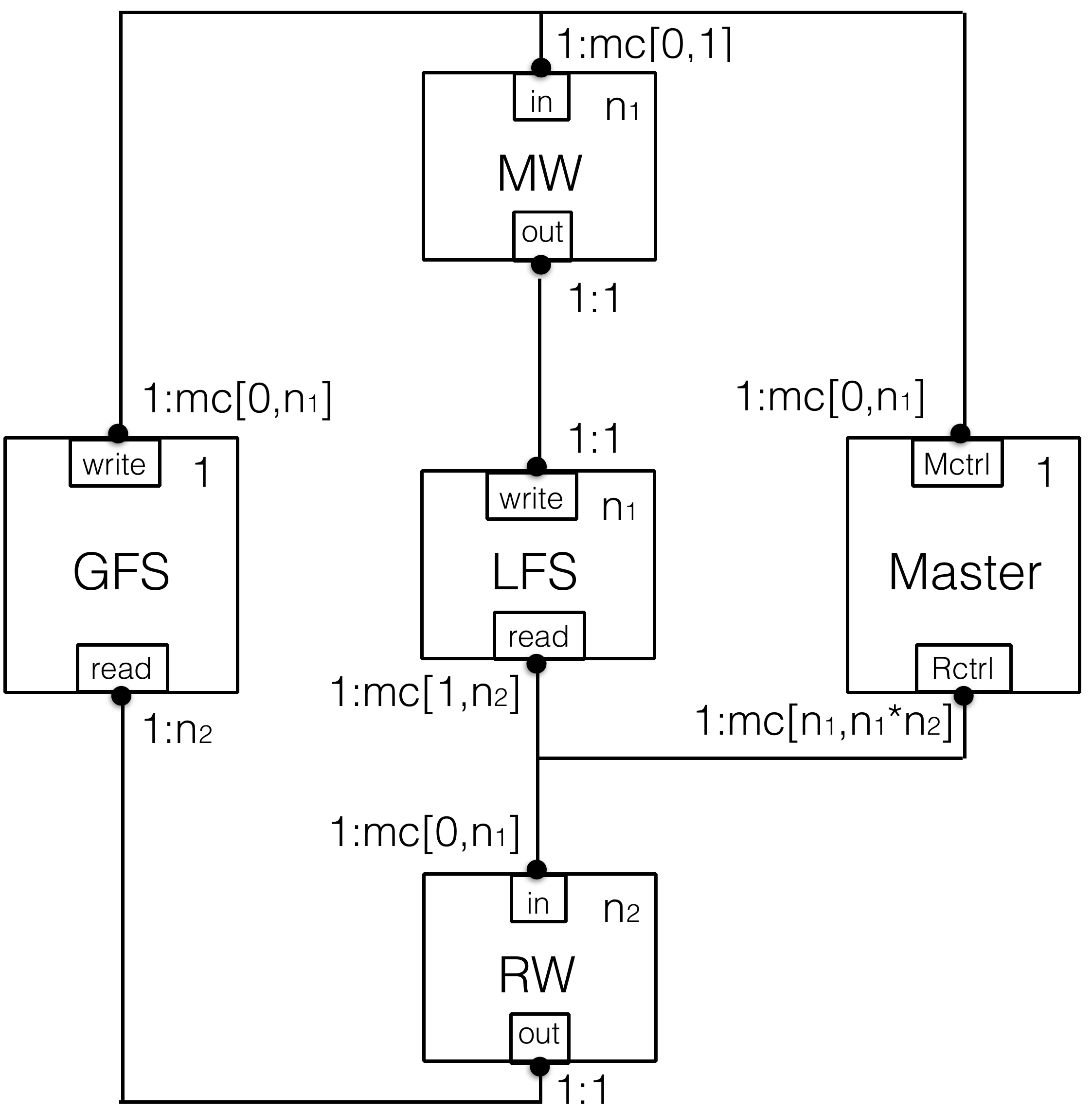}
	\caption{Map Reduce architecture style.}  	
  	\label{fig:mapsstyle}
\end{minipage}
\begin{minipage}[b]{0.5\textwidth}
	\centering
	\includegraphics[scale=0.25]{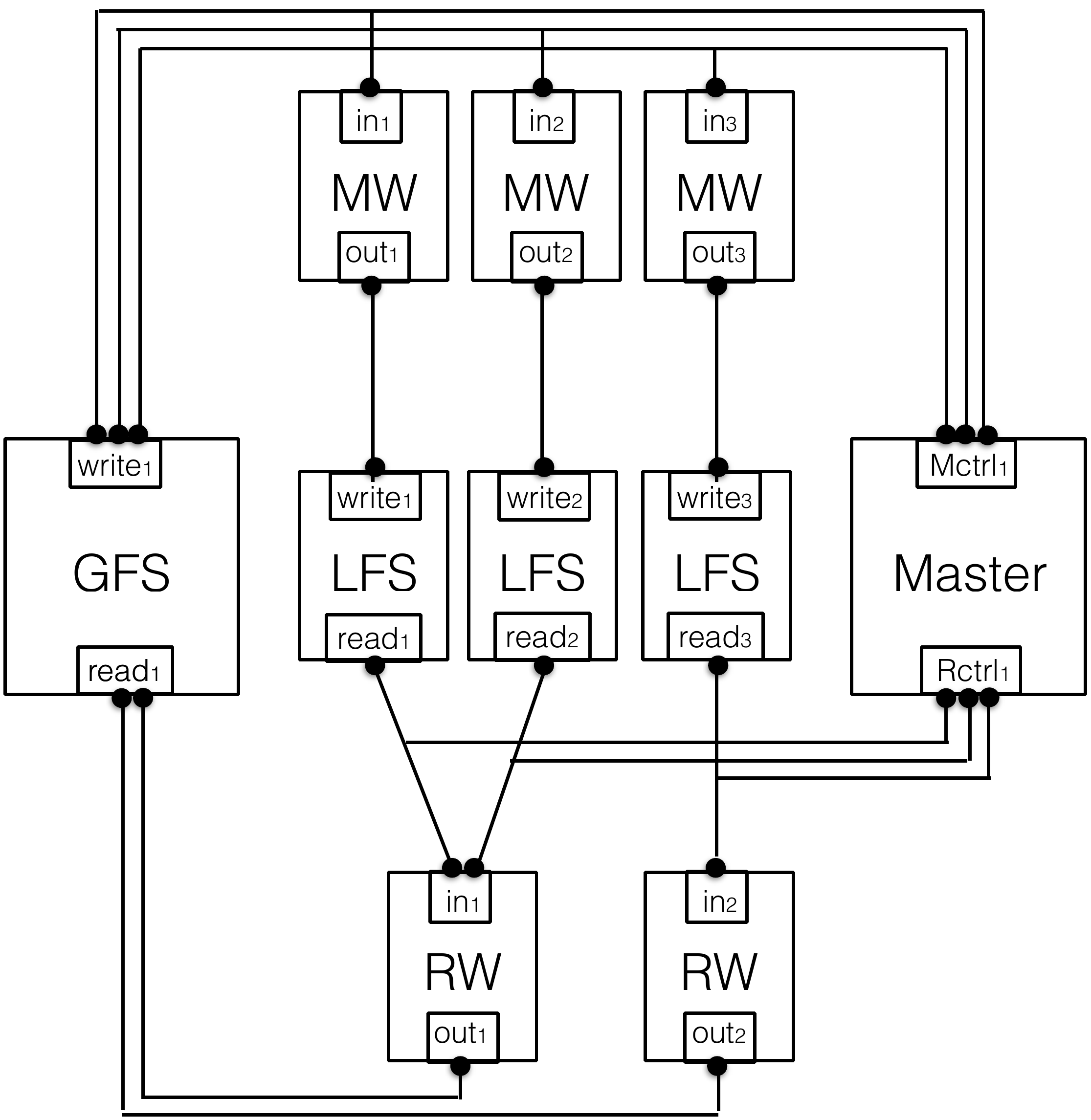}
  	\caption{Map Reduce architecture.}
  	\label{fig:mapsconf}
\end{minipage}
\end{figure}

A large dataset is split into smaller datasets and stored in the global filesystem ($GFS$).  The $Master$ is responsible for coordinating and distributing the smaller datasets from the $GFS$ to each of the map workers ($MW$).  The port $in$ of each $MW$ is connected to the $Mcontrol$ and $read$ ports of the $Master$ and the $GFS$, respectively.  Each $MW$ processes the datasets and writes the result to its dedicated local filesystem ($LFS$) through a binary connector between their $out$ and $write$ ports.  The connector is binary since no $MW$ is allowed to read the output of another $MW$.  Each reduce worker ($RW$) reads the results from multiple $LFS$ as instructed by the $Master$.  To this end, the $in$ port of each $RW$ is connected to the $Rcontrol$ and $read$ ports of the $Master$ and some $LFS$, respectively.  Each $RW$ combines the results and writes them back to the $GFS$ through a binary connector between their $out$ and $write$ ports. 
\end{example}

\vspace{-0.5cm}

\section{Checking Conformance}
\label{sec:satisfaction}

Algorithm~\ref{alg:verifymodel} with polynomial-time complexity checks whether an architecture $\langle \cB, \gamma \rangle$ conforms to a simple diagram $\langle \cT, n,  \cC \rangle$.  It can be easily extended for interval diagrams as shown in~\cite{MBBS16-Diagrams-TR}.

Algorithm \ref{alg:verifymodel} checks the validity of the following three statements: 1) the number of components of each type $T$ is equal to $n(T)$;
 2) there exists a partition of $\gamma$ into $\gamma_1, \dots, \gamma_l$ such that each $\gamma_i$ corresponds to a different connector-motif $\cm_i \in \cC$ of the diagram; 3) for each connector motif $\cm_i$ and its corresponding $\gamma_i$, the number of times each port instance participates in $\gamma_i$ satisfies the degree constraints. The three statements correspond to functions \ref{alg:verifycardinality}, \ref{alg:verifymultiplicity} and \ref{alg:verifydegree}, respectively. If all statements are valid the algorithm returns \emph{true}, \ie the architecture conforms to the diagram.

In particular, function \ref{alg:verifycardinality} takes as input the architecture diagram $\langle \cT, n,  \cC \rangle$ and the set of components $\cB$ of the architecture $\langle \cB, \gamma \rangle$. It counts the number of components for each component type in $\cB$ and it returns \emph{true} if for each component type $\cT$ of the diagram its cardinality matches the corresponding number of components in $\cB$. Otherwise it returns \emph{false} and algorithm \ref{alg:verifymodel} terminates.


Function \ref{alg:verifymultiplicity} takes as input the configuration $\gamma$ of the architecture $\langle \cB, \gamma \rangle$ and the set of connector motifs $\cC$ of the architecture diagram $\langle \cB, \gamma \rangle$. The function checks whether there exists a partition of $\gamma$ such that each sub-configuration $\gamma_i$ of $\gamma$ corresponds to a distinct connector motif $\cC_i$ of $\cC$, \ie each connector $k$ in $\gamma_i$ conforms to the multiplicity constraints of $\cC_i$. If such a partition exists the function returns it. Otherwise, it returns $\emptyset$ and algorithm \ref{alg:verifymodel} terminates.



 Function \ref{alg:verifydegree} takes a connector motif $\Gamma$ of $\cC$ and its corresponding sub-configuration of $\gamma$ assigned by \ref{alg:verifymultiplicity}. For each port instance in the sub-configuration it checks whether the number of times the port participates in different connectors is equal to the corresponding degree constraint of the connector motif. If the check fails, algorithm \ref{alg:verifymodel} terminates.

Algorithm~\ref{alg:verifymodel} uses a number of auxiliary functions.
Function \texttt{generic(p)} takes a port instance and returns the corresponding generic port. Function \texttt{typeof(B)} returns the component type of component \texttt{B}. Operation \texttt{map[key]++} increases the \texttt{value} associated with the \texttt{key} by one if the \texttt{key} is in the \texttt{map}, otherwise it adds a new \texttt{key} with \texttt{value} $1$.

\begin{figure} [t]
  \begin{minipage}[t]{0.49\textwidth}
    \begin{algorithm}[H]
\label{algo_verify_model} 
\KwData{Architecture $\langle \cB, \gamma \rangle$, diagram $\langle \cT, n, \cC \rangle$}

\KwResult{Returns $true$ if the architecture satisfies the diagram $\langle \cT, n,  \cC \rangle$. Otherwise returns $false$.}

\BlankLine
\SetKw{KwNot}{not}
\If{ \KwNot VerifyCardinality($\cB$, $\langle \cT, n,  \cC \rangle$)}{
		\Return $false$\;
}
\BlankLine
$\mathcal S_{\gamma} \longleftarrow $ \textit{VerifyMultiplicity($\gamma$, $\cC$)}\;

\If{$\mathcal S_{\gamma} = \emptyset$} {
	\Return $false$\;
}
\mbox{}\\[1pt]
\For{$\cm \in \cC$}{
	\If{not VerifyDegree($\mathcal S_{\gamma}[\cm], \cm$)}{
		\Return $false$\;		
	}	
} 
\Return $true$\;


\caption{VerifyArchitecture}
\label{alg:verifymodel}
\end{algorithm}
\end{minipage}
\hspace{2pt}
\begin{minipage}[t]{0.49\textwidth}
\begin{function}[H]

\KwData{Set of components $\cB$, diagram $\langle \cT, n,  \cC \rangle$}

\KwResult{Returns $true$ if the number of components of each type in $\cB$ is equal to corresponding cardinality of the diagram. Otherwise, it returns $false$. }

\BlankLine
\tcc{Map with key: type, value: number of instances} 
$countTypes \longleftarrow \{\}$\;

\For{$B_i \in \cB$}{
		$countTypes[typeof(B_i)] ++$\;
}

\For{$T_i \in \mathcal T$}{
	\If{$countTypes[T] \neq n[T]$}{
		\Return $false$\;		
	}	

}
\Return $true$\;

\caption{VerifyCardinality($\cB$, $\langle \cT, n,  \cC \rangle$)}
\label{alg:verifycardinality} 
\end{function}
\end{minipage}
\end{figure}

\newlength{\templng}

\vspace{-0.4cm}

\section{Related Work}
\label{sec:relatedwork}

A plethora of approaches exist for architecture specification.
Patterns~\cite{daigneau2011service,Hohpe:2003:EIP:940308} are commonly used for specifying architectures in practical applications. The specification of architectures is usually done in a graphical way using general purpose graphical tools. Such specifications are easy to produce but the meaning of the design may not be clear since the graphical conventions lack formal semantics and thus, are not amenable to formal analysis. 

A number of Architecture Description Languages (ADLs) have been developed for architecture specification~\cite{medvidovic2000classification, woods2005architecture, ozkaya2013we}.  
Nevertheless, according to~\cite{malavolta2013industry}, architectural languages used in practice mostly originate from industrial development instead of academic research. Practitioners insist on using UML even though it lacks formal semantics. ADLs with formal semantics require the use of formal languages which are considered as difficult for practitioners to master~\cite{malavolta2013industry}. 
To address this issue, we propose architecture diagrams that combine the benefits of graphical languages and rigorous formal semantics. By relying on the minimal set of notions, we emphasize the conceptual clarity of our approach. 

Architecture diagrams were developed to accommodate architecture
specification in BIP~\cite{AttieBBJS15-architectures-faoc}, wherein
connectors are $n$-ary relations among ports and do not carry any
additional behaviour.  This strict separation of computation from
coordination allows reasoning about the coordination constraints
structurally and independently from the behaviour of coordinating
components.  However, our approach can be extended to describe
architecture styles in other coordination languages by explicitly
associating the required behaviour to connector motifs.  In
particular, this can be applied to specify connector patterns in
Reo~\cite{reo}, by associating multiplicity and degree to source and
sink nodes of connectors.
The main difficulty is to correctly instantiate the behaviour depending on the number of ends in the connector. 

\begin{wrapfigure}{l}{0.5\textwidth}
\begin{function}[H]
\settowidth{\templng}{Map with key: $\cm$, value: sub-configuration blah blah blah }
\KwData{Configuration $\gamma$, set of connector motifs $\cC$}

\KwResult{Returns a partition $\mathcal P_{\gamma}$ of $\gamma$ into connectors that satisfy the multiplicity constraint. If no partition exists, it returns $\emptyset$. }

\BlankLine
\tcp{\makebox[\templng]{Map with key: $\cm$, value: sub-configuration\hfill}} 
$partition \longleftarrow \{\}$ 

\For{$\cm \in \cC$}{
	$partition[\cm] \longleftarrow \emptyset$;
}
\BlankLine

\tcp{Map with key: generic port, value: number of port instances in connector}
\For{$k \in \gamma$}{
	$portscount \longleftarrow \{\}$ 
	\For{$p_i \in k$}{
		$portscount[generic(p)] ++$\;		
	}
	$x \longleftarrow false$\;
	\For{$\cm = (a, \{m_p:d_p\}_{p \in a}) \in \cC$}{
 		\If{$a = keys(portscount)$}{
 			$y \longleftarrow true$\;
 			\For{$p \in a$}{
				\If{$portscount[p] \neq m_p$}{
					$y \longleftarrow false$\; \textbf{break}\;	
				}				
 			}
 			\If{$y$}{
				$partition[\cm] \longleftarrow partition[\cm] \cup k$\;
				$x \longleftarrow true$\; \textbf{break}\;	
 			}
		}	 		
     }

     \If{$x = false$}{
     	\Return $\emptyset$\;
     } 
}
\Return partition\;

\caption{VerifyMultiplicity($\gamma$, $\cC$)}
\label{alg:verifymultiplicity} 
\end{function}
%

\begin{function}[H]
\settowidth{\templng}{Map with key: port, value: number of connectors blah blah blah}
\KwData{Configuration $\gamma_i$, connector motif $\cm$
}

\KwResult{Returns $true$ if the degree requirements are satisfied. Otherwise, it returns false. }


\BlankLine
 \tcp{{Map with key: port, value: number of connectors}}
$degrees \longleftarrow \{\}$   

\For{$k \in \gamma_i$}{ 
	\For{$p_i \in k$}{
		$degrees[p_i] ++$\;   		
	}
}

\For{$p_i \in keys(degrees)$}{ 
	\If{$degrees[p_i] \neq d_{generic(p_i)}$}{
		\Return $false$\; 
	}
}
\Return $true$\;

\caption{VerifyDegree($\gamma_i$, $\cm$)}
\label{alg:verifydegree} 
\end{function}
\end{wrapfigure}

Alloy~\cite{alloy} has been used for architecture style specification, in the ACME~\cite{kim2010analyzing} and
Darwin~\cite{darwin} ADLs. The connectivity primitives in~\cite{kim2010analyzing,darwin}
are binary predicates and cannot tightly characterize coordination
structures involving multiparty interaction.  To specify an $n$-ary
interaction, these approaches require an additional entity connected
by $n$ binary links with the interacting ports.  Since the behaviour
of such entities is not part of the architecture style, it is
impossible to distinguish, \eg between an $n$-ary
synchronisation and a sequence of $n$ binary ones.

 Architecture diagrams consist of component types and connector motifs, respectively comparable to UML components and associations~\cite{ivers2004documenting, umlSpecs}. One important difference between connector motifs and UML associations is that the latter cannot specify interactions that involve two or more instances of the same component type~\cite{umlSpecs}. In UML, the term ``multiplicity'' is used to define both 1)~the number of instances of a UML component and 2)~the number of UML links connected to a UML component. 
In architecture diagrams, we call these, respectively, ``cardinality'' and ``degree''.  We use the term ``multiplicity'' to denote the number of components of the same class that can be connected by the same connector.  The distinction between multiplicity and degree is key for allowing $n$-ary connectors involving several instances of the same component type.

A large body of literature, originating in
\cite{montanari99-graphs,metayer98}, studies the use of graph grammars
and transformations~\cite{rozenberg1997handbook} to define software
architectures.  Although this work focuses mainly on dynamic
reconfiguration of architectures, \eg
\cite{adr,koehler2008connector,krause2011modeling}, graph grammars can
be used to define architecture styles: a style admits
all the configurations that can be derived by its defining grammar.  
The use of context-free grammars allows inductive
definitions and reasoning about architectures.  The downside is that
such definitions require additional non-terminal symbols to represent
variable size structures, \eg list of all slaves in a Master/Slave
architecture.
We take a different approach, whereby all constraints appear directly
in the architecture diagram for which we provide denotational
semantics.  The rationale is the following: we
assume that the reasoning is carried out by an ``expert'', who defines
the architectural style, whereas the ``user'' only needs the minimal
information in order to select and instantiate it.  Thus, structural
information, \eg necessary information for an inductive proof that the style
imposes a certain property, does not appear in the diagram, but only
the entities that form the target system.

\vspace{-0.4cm}
\section{Conclusion and Future Work}
\label{sec:discussion}

We studied architecture diagrams, a graphical language rooted in well-defined semantics for the description of architecture styles. We studied two classes of diagrams. Simple architecture diagrams express uniform degree and multiplicity constraints. They are easy to interpret and use but have limited expressive power. Interval architecture diagrams allow heterogeneity of multiplicity and degree and thus, are strictly more expressive. Architecture diagrams provide powerful and flexible means for graphical specification of architectures with $n$-ary connectors.
Using architecture diagrams instead of purely logic-based specifications confers the advantages of graphical formalisms.  

In an ongoing project partially financed by the European Space Agency, we are using architecture diagrams to describe architectures in the case studies of the project. 
 We are currently working on extending the current notation with arithmetic constraints and implementing the synthesis procedure described in this paper with the JaCoP\footnote{http://jacop.osolpro.com/} constraint solver. In the future, we plan to extend connector motifs with data flow information and study the expressive power of architecture diagrams.
\vspace{-0.4cm}
\bibliographystyle{eptcs}
\bibliography{cl}

\end{document}